# Atomically Thin Yellow Pearl: An Impetus for Thermal Nonlinear Optical Effect Assisted Continuous Wave Random Lasing

Nabarun Mandal[1], Ashim Pramanik[2], Arindam Dey[3], Pathik Kumbhakar[3], Vidya Kochat[4*], Abhay Raj Singh Gautam[5], Nicholas Glavin[6], Ajit K Roy[6], P. M. Ajayan[7], and Chandra Sekhar Tiwary[2*]

[1]*School of Nanoscience and Technology, Indian Institute of Technology Kharagpur, 721302, India.*

[2]*Department of Metallurgical and Materials Engineering, Indian Institute of Technology Kharagpur, 721302, India.*

[3]*Nanoscience Laboratory, Department of Physics, National Institute of Technology, Durgapur, 713209, West Bengal, India.*

[4]*Materials Science Centre, Indian Institute of Technology Kharagpur, 721302, India.*

[5]*Materials Engineering, Indian Institute of Technology, Gandhinagar 382355, India.*

[6]*Materials and Manufacturing Directorate, Air Force Research Laboratory, Wright Patterson AFB, Ohio 45433-7718, United States.*

[7]*Department of Materials Science and Nano Engineering, Rice University, Houston, Texas 77005, United States.*

*\*Corresponding authors, E-mail: chandra.tiwary@metal.iitkgp.ac.in, vidya@matsc.iitkgp.ac.in.*

## ABSTRACT

Above cryogenic temperature, random laser (RL) emission under continuous wave (CW) optical pumping is often challenging for materials having plenty of lattice defects. Thus, intensive care is required while fabricating the RL device. Synthesis of 2D materials having unique functionality at the atomic limit is necessary for use in such devices. In this work, defects-free 2D yellow pearl (2D-YP) has been synthesized from bulk south sea pearl using liquid-phase exfoliation (LPE) technique. Thereafter, 2D-YP has been employed as passive scatterer to achieve CW pumped RL emission from a conventional gain molecule, *i.e.*, Rhodamine B (RhB). Compared to other semiconductor ($TiO_2$) and 2D (Graphene and h-BN) passive scatterers, ~25 times improvement in gain volume is observed for the disordered system consisting of RhB and 2D-YP, *i.e.*, 2D-YP@RhB, which is found to be due to the formation of a large refractive index gradient, as induced by the thermal nonlinear optical (NLO) interaction. Hence, incoherent RL (ic-RL) emission ~591 nm is reported for 2D-

YP@RhB at a lowest threshold input pump power and a linewidth of 65 mW and 3 nm, respectively. Additionaly, a clear transition from ic-RL to mode tunable coherent RL (c-RL) emission is also possible by altering the configuration of the 2D-YP@RhB container. Therefore, the newly designed Van der Waals heterostructure, *i.e.*, 2D-YP with intrinsic photon trapping capability may pave an avenue towards developing future exciting optoelectronic devices.

**Keywords:** 2D yellow Pearl (2D-YP); Liquid Phase exfoliation (LPE); optical properties; random laser.

The emergence of 2D materials, like graphene and beyond, has attracted attention from the photonics community due to their interesting electronic and optical properties. [1-5] Quantum confinement of electrons in the direction perpendicular to the surface of the 2D material leads to size and surface dependent optical properties that is absent in its bulk form. The presence of strong light-matter interaction in 2D materials makes them suitable for a wide range of photonic applications. [6-8] For instance, Liu *et al.* [3] fabricated a mode locked laser by regulating defects in 2D $Bi_2O_2Se$. Designing of random metamaterial using graphene for cavity free stimulated emission, *i.e.*, random laser, has been also demonstrated. [9] Generation of random laser (RL) emission, by modulating the propagation and amplification of photons inside a disordered active medium, has created a lot of research interest in photonics community. [10-13] Unlike the RL emission under pulse laser pumping, generation of RL emission at room temperature and continuous wave (CW) optical pumping is sometimes difficult to achieve due to a number of potential causes such as simple heating, formation of photo-generated quenchers, and dielectric screening. [14, 15] Notably, lattice defect in 2D material may act as a barrier for demonstrating room temperature RL emission under CW pumping. Till now, the hurdles in generating CW pumped RL emission has been overcome by means of sophisticated device architecture, [16] and use of single-crystalline 2D perovskites having low defect density. [17, 18] However, both fabrication CW RL devices and synthesis of low defect 2D perovskites requires costly instrumental facilities as well as several optimization process.

Therefore, people are now looking for alternative light scattering 2D materials which can localize photons inside a disordered system through unconventional processes of feedback for RL emission under CW and pulsed laser pumping. [19, 20] In this regard, trapping of photons inside gain medium can be done by utilizing the nonlinear optical response (NLO) of the disordered system in presence of a 2D material as passive scatterer. [9, 21] Also, thermal NLO effect offer the change of *in-situ* temperature dependent refractive gradient inside a liquid disordered medium consisting of 2D materials. Therefore, the objective of RL emission just by utilizing the refractive index contrast within the active medium can be fulfilled easily, as already reported for semiconductor nanostructures under pulse optical pumping. [22, 23] Recently, the focus of many materials engineers has shifted towards naturally available materials to exfoliate into 2D depending upon their availability and procurement costs. Naturally available bulk materials with laminar structures are suitable for liquid phase exfoliation. [24, 25] In this regard, ultrasonication based liquid phase exfoliation (LPE) technique is preferable to shear-induced or electrochemical LPE due to highly homogeneous dispersions of 2D materials being obtainable by this process. [26-28] During the last decade, significant efforts to obtain nanomaterials using LPE has resulted in the production of wide number of conventional 2D materials in different solvents. [29-32] Interestingly, exfoliation of natural van der Waals heterostructures from bulk franckeite was demonstrated by Li *et al.* [24] Mahapatra *et al.* have synthesized atomically thin biotene by exfoliating naturally abundant biotite. [25] Another such naturally occurring material with a laminar structure is pearl. A pearl is a hard, iridescent, circular object produced by a living shelled mollusc within its mantle (soft tissue) when an invasion by a foreign object threatens to harm the creature. Pearls are quite unique because they are organic gemstones and aren't considered to be true minerals. The basic structure of a pearl is composed of concentric layers of calcium carbonate in the crystalline form aragonite, or sometimes a mixture of aragonite and calcite, held together by a naturally produced organic compound called conchiolin. This structure of thickness 300-1500 nm is called nacre or mother-of-pearl. [33] Under a microscope, this structure resembles a terrace-like formation stacked together in a brick wall fashion (sheet nacre). [34] The most valuable of the cultured pearls is the south sea pearl, known for its golden hue and produced by the pearl oyster of the species *Pinctada maxima*. [35] Additionally, the bulk pearl is already known to have interesting optical properties. [36]

Here, 2D yellow pearl (2D-YP) has been synthesized from its bulk form by using a simple liquid phase exfoliation (LPE) technique. The structural and linear optical properties of the newly synthesized 2D material is analyzed through standard characterization tools and light localization capability of 2D-YP is experimentally demonstrated. The presence of 2D-YP along with conventional dye molecules, *i.e.*, Rhodamine B (RhB), enhances the thermal nonlinear optical (NLO) interaction in order to create a temperature dependent refractive index gradient. Coupling the advantageous optical properties of 2D-YP, RL emission is achieved from the disordered system 2D-YP@RhB which contains RhB dye as gain molecule and 2D-YP as passive light scatterer. Also, performance of 2D-YP for generating RL emission is compared to other semiconductors and 2D materials. In diffusive regime, formation of largest gain volume in 2D-YP@RhB due to the highest value of thermo-optic constant ($dn/dT$) results ic-RL emission nm at a lowest threshold pump power ($^{Th}P_{in}$) of 65 mW. Moreover, transformation of incoherent RL (ic-RL) emission at ~ 591 nm to coherent (c-RL) emissions at ~587 and 589 nm is observed when 2D-YP@RhB is placed inside a 5 mm cuvette and capillary tube of inner diameter 1 mm, respectively. Such a transformation from ic-RL to c-RL emission is due to enhanced possibility of multiple scattering, as confirmed in this work through a phenomenological experiment. Altogether, this study demonstrates a simple approach for synthesizing unconventional 2D materials from natural resources for promising photonic applications *i.e.*, designing of RLs and spatial self-phase modulators.

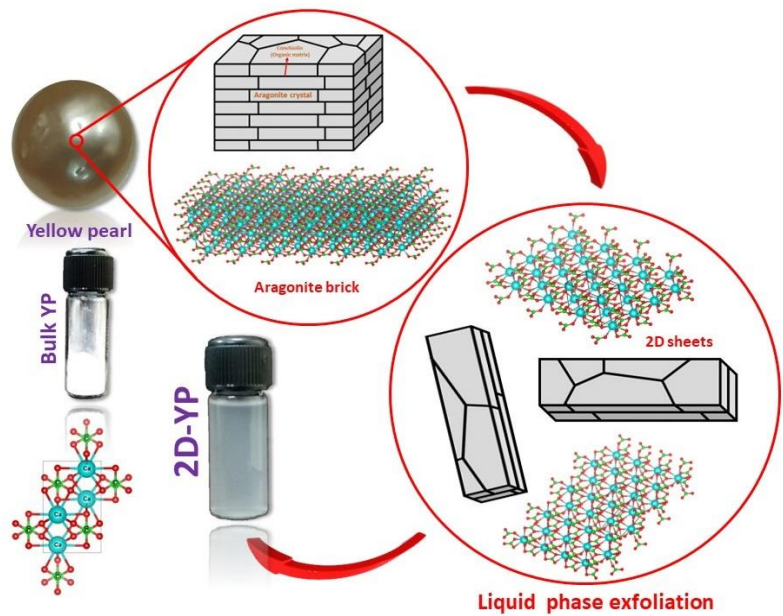

**Figure 1.** Schematic representing the synthesis of 2D-YP from bulk yellow south sea pearl.

## RESULTS AND DISCUSSIONS

### Structural and Optical Properties of 2D-YP

A simple LPE process is opted for synthesizing the 2D-YP from bulk south sea yellow as shown schematically in **Figure 1**. HRTEM images as shown in **Figure 2**a-b shows atomic structure and surface profile plots revealing an arrangement of Ca and C atoms. **Figure 2**a shows 2D sheet formation in the form of a few overlapping layers. The inverse Fast Fourier transform (FFT) lattice fringes with a d-spacing value of 0.276 nm match well with the (121) plane of aragonite. This plane is noted to have high intensity from XRD data due to its maximum exfoliation in **Figure 2**c. HRTEM pattern shows atomic-level hexagonal crystal structures in **Figure 2**b. To understand the atomic arrangement, surface line plots marked line A and line B were traced. Profile A shows Ca atoms at higher intensities along with C atoms and O vacancies at lower intensities while profile B shows Ca atoms with C atoms placed between. This reveals the orthorhombic crystal structure of aragonite. [37] The crystalline part of exfoliated 2D-YP sample is observed to be made up of aragonite only with the other polymorphs of $CaCO_3$ absent.

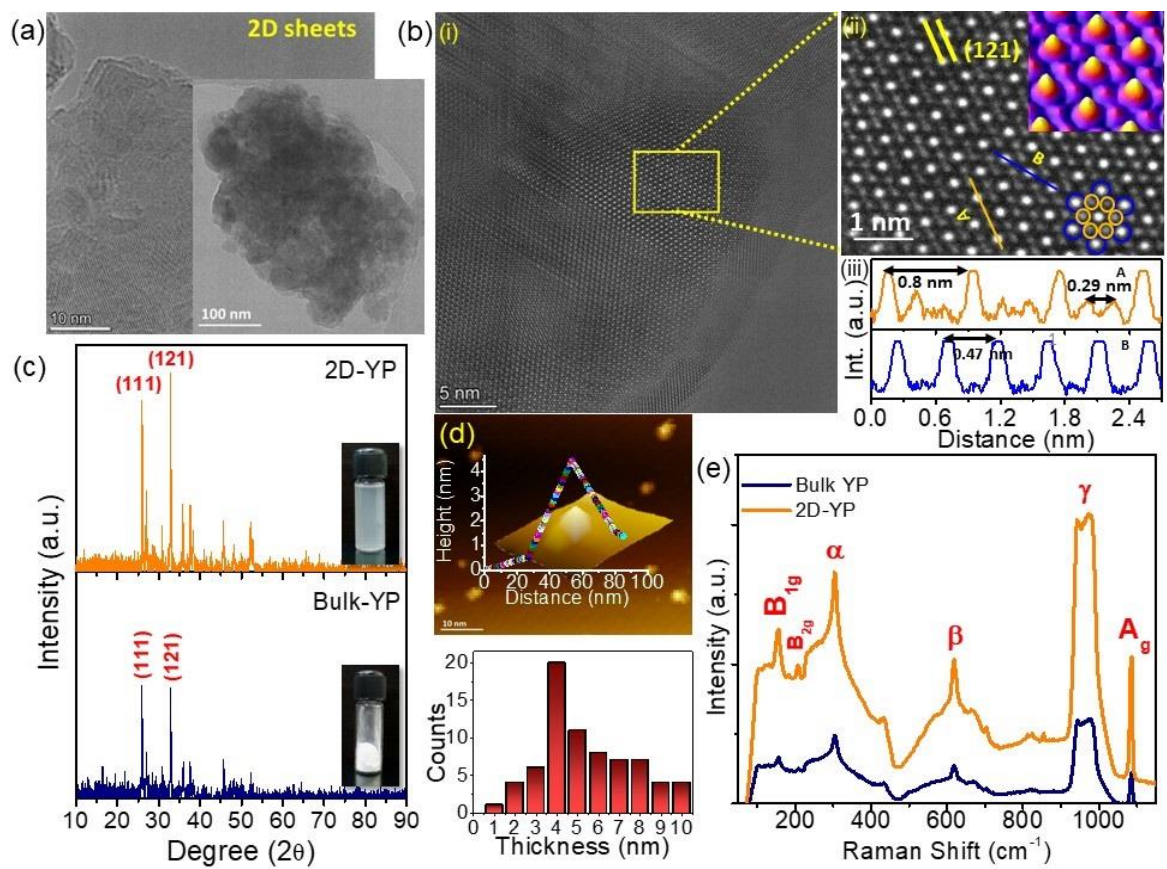

**Figure 2.** a), b) (i-iii) HRTEM images of 2D-YP sample sheet formation, $CaCO_3$ crystal structure and surface line profiles showing a d-spacing of 0.276 nm. c) XRD data of bulk YP and exfoliated 2D-YP samples. d) AFM image (i) and the analysis revealing material thickness (ii). e) Raman spectra for bulk YP and 2D-YP samples (α= 304 cm$^{-1}$, β= 618 cm$^{-1}$ and γ= 975 cm$^{-1}$).

XRD spectra of both bulk yellow pearl nacre and the exfoliated 2D-YP sample are shown in **Figure 2**c. The obtained bulk spectrum is consistent with that of aragonite and matches the orthorhombic structure with reference to ICDD: 00-005-0453 and space group Pmcn, 62. [38] Usually, the (111) crystalline plane of aragonite can be identified as the strongest standard diffraction peak in XRD spectra of nacre, as obtained in bulk. But in **Figure 2**c, the strongest XRD diffraction peak obtained is due to the (121) crystalline plane of aragonite. This plane of aragonite is interesting because the C-crystal axis of (121) crystalline plane is oblique to the stratification plane of the nacre. [38] The obtained XRD spectra thus show that the (121) crystalline plane of aragonite has strong preferential orientation in 2D-YP sample. In order to investigate the thickness of 2D sheets in synthesized sample, AFM analysis was carried out. The thickness of the 2D material was calculated using line profile and a histogram was obtained as shown in **Figure 2**d. The material thickness was ~4-5 nm and the lateral width was in the range of 100-400 nm. Raman spectroscopy is an important tool to identify electron-phonon coupling, symmetric stretching and layer thickness of 2D materials. Aragonite, main component of nacre, has 30 active peaks in Raman. These peaks are theoretical and not all of them will be detectable due to the presence of background noise, low peak intensities and limitations in resolution of the instrument. [39] The Raman spectrum of 2D-YP in comparison with its bulk form is shown in **Figure 2**e. The peak at 1085 cm$^{-1}$ corresponds to the symmetric stretching mode ($v_1$) of the carbonate ion. In the low energy region of the spectrum, lattice vibration modes were identified in the form of scattering signals at 152 cm$^{-1}$ and 206 cm$^{-1}$. The peaks at 152 cm$^{-1}$, 206 cm$^{-1}$ and 1085 cm$^{-1}$ corresponds to the $B_{1g}$, $B_{2g}$ and $A_g$ symmetries, respectively, of aragonite crystal. [39] The peaks at 304 cm$^{-1}$, 618 cm$^{-1}$ and 975 cm$^{-1}$ are due to the complex organic compound conchiolin.

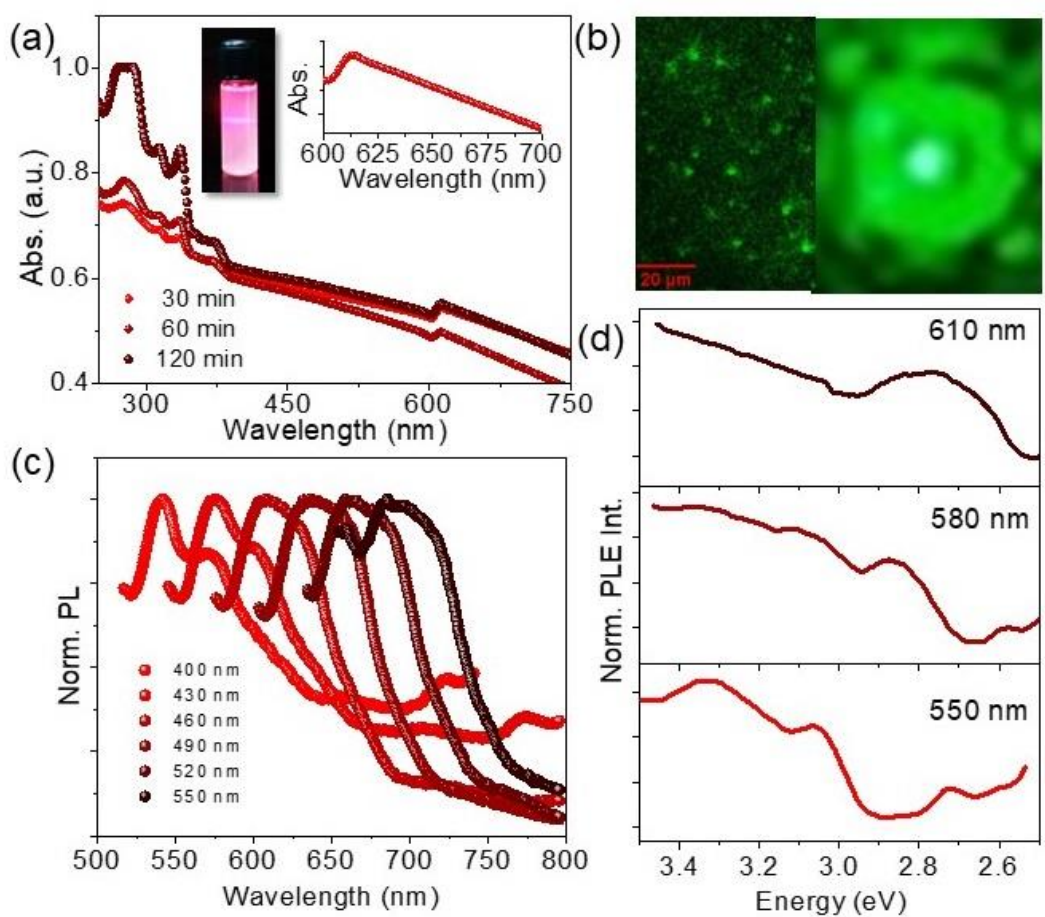

**Figure 3.** a) UV-Vis absorbance spectrum of 2D-YP along with a demonstration of Tyndall effect. b) Light localization images of exfoliated 2D-YP taken under 532 nm laser. c) PL at various excitation wavelengths for 2D-YP. d) PLE at various emission wavelengths for 2D-YP.

For the optical characterization of 2D-YP, absorbance spectra of 30 min, 60 min and 120 min exfoliated samples has been obtained at room temperature, as shown in **Figure 3**a. A characteristic peak at 277 nm was obtained for all time-variant exfoliated samples. The intensity of this peak increases significantly after 60 min of probe sonication signalling material exfoliation as the organic matrix surface gets more exposed. This peak isn't due to the pearl surface structure or aragonite crystal structure but is characteristic of the organic matrix conchiolin. [36] The absorbance band stays flat at 400-600 nm. Additionally, a smaller peak is seen at 617 nm. This peak is found in saltwater cultured pearls and is due to the presence of colour pigments in south sea yellow pearl (porphyrins). [40] A simple Tyndall effect depicting dispersed 2D-YP sample scattering light under 632 nm laser excitation is shown in the inset of **Figure 3**a. In order to observe localization and scattering of light, the material dispersion was placed under a 532 nm laser and optical imaging was done. The exfoliated sample has several localized points of light appearing as light halos (**Figure 3**b), which is similar to quasicrystals. [41]

To understand the light emission characteristics of the synthesized 2D-YP, excitation wavelength dependent photoluminescence (PL) spectra of the synthesized sample was obtained and is shown in **Figure 3**c. At low excitation wavelength of 400 nm, two emission peaks are obtained at ~540 and 725 nm. These two PL signals are related to electronic transitions from two different energy states. However, the peak at ~725 nm disappears at higher excitation wavelengths. Also, the peak at ~540 nm shifts to higher wavelengths and broadens as the excitation wavelength is increased. To further identify the energy states, emission wavelength dependent PL excitation (PLE) spectra was also obtained as shown in **Figure 3**d. The PL emission at 550 nm is corresponding to the electronic transition from the energy bands at ~ 2.72, 3.00 and 3.33 eV. Whereas, PL emission at longer wavelengths originates from the energy bands ~2.70 to 2.90 eV.

**Thermal NLO effect induced RL Generation**

Based on their fascinating optical characteristics, 2D-YP has been employed as a passive scatterer to generate random laser (RL) emission from Rhodamine-B dye (1 mM, gain medium). A commonly used experimental set up is used for RL generation experiment. [42] The schematic of the RL experimental set up used for measurement of emission characteristics is shown in **Figure 4**a. Briefly, a 100 mW CW diode laser of wavelength 532 nm has been used for optical pumping and is focused onto the surface of quartz cuvette by using a convex lens of focal length 10 cm. A fiber optic detector (Thorlabs) has been used to detect the emission spectra, and the RL emission is captured and analyzed by Thorlabs OSA software in a computer.

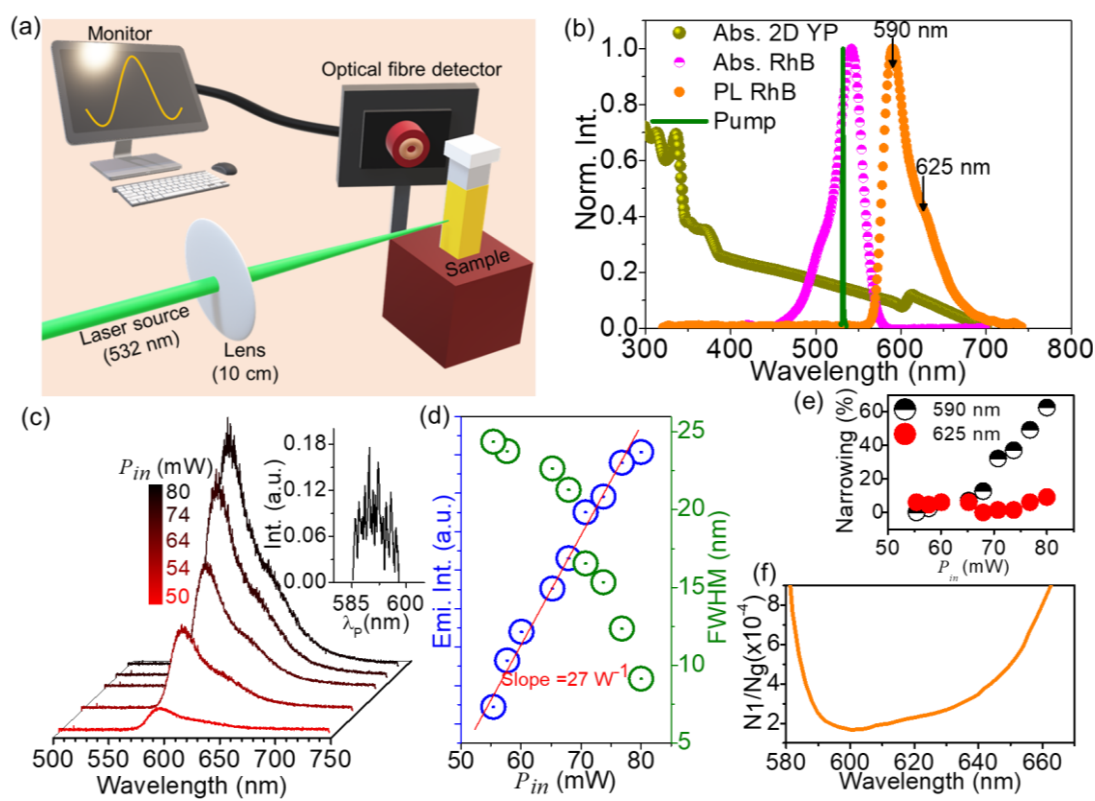

**Figure 4.** a) Experimental set up for RL generation in 2D-YP@RhB. b) UV-Vis absorption and PL emission ($\lambda_{ex}$=532 nm) spectra of the 2D-YP and RhB dye. c) Emission spectrum of bare RhB dye (1 mM) at different pump powers. d) Variation of output emission intensity and FWHM with $P_{in}$. e) Variation FWHM of emission peak of the system 2D-YP@RhB at ~ 590 and 625 nm with $P_{in}$. f) The fraction of RhB molecules in the excited state ($N_1/N_g$) at different emission wavelengths as estimated from **Equation 1**.

However, prior to doing the RL experiment we have done a comparative study of optical characteristics of the scatterer, gain medium and pump laser as shown in **Figure 4**b. The performance of a luminescent species in RLs as a gain medium depends on the separation between its absorption and emission bands, i.e., emission overlap (EO) integral. A low value of EO signifies the least possibility of photo reabsorption within the gain medium. Herein, we have estimated the value of EO integral for the gain medium [43], and it is found to be ~0.25. Meanwhile, the small absorption of 2D-YP at ~532 nm, allowed RhB dye to absorb the most of the incident radiation without any extra optical loss. The scattering property of 2D-YP may also arise due to the contrast in the refractive index [44-46] as confirmed later using NLO experiments. We have also tested the stability of the RhB dye by continuously exposing it under 532 nm CW laser for 5 min (**Supporting Figure S1**)

At first, the emission characteristics of the bare RhB dye solution is studied at different pump powers ($P_{in}$) as shown in **Figure 4c** and **4**d. Discrete sharp peaks are found to be absent on the top of the spontaneous emission background for the bare dye solution, when the pump power $P_{in}$ is increased from 50 to 80 mW. The emission behavior of bare RhB is observed to be more like amplified spontaneous emission (ASE). Also, **Figure 4**d depicts the variation of emission intensity of the spectrum shown in **Figure 4**c as a function of $P_{in}$. Clearly, peak emission intensity increases almost linearly with the increase of the $P_{in}$ with a slope of 27 W$^{-1}$. Notably, the emission spectra of RhB dye consist of one sharp peak at ~590 nm with a shoulder emission center at ~625 nm (**Figure 4**b). Spectral narrowing upto 60% is observed for the peak ~590 nm on increasing the $P_{in}$. Whereas, FWHM of the peak at ~590 nm remains almost constant with $P_{in}$, as can be seen from **Figure 4**e. This further confirms the ASE behavior of bare RhB solution. [47, 48] Moreover, to find out the origin of emission wavelength selective spectral narrowing behavior in RhB dye, we have estimated the fraction of gain molecules ($N_1/N_g$) in the excited state as a function of emission wavelength ($\lambda$) by using the following relation; [49]

$$\frac{N_1}{N_g} = \frac{1}{(\sigma_e(\lambda) + \sigma_a(\lambda))} \left[ \sigma_a(\lambda) + \frac{2\pi m}{N_g \lambda \times 10^6} \right] \qquad (1)$$

Where, $\sigma_e(\lambda)$, $\sigma_a(\lambda)$ and $m$ are the emission cross section, absorption cross section and refractive index of the gain medium, respectively. $N_g = 1.20 \times 10^{17}$ nos. ml$^{-1}$ is the number density of the gain molecules. $N_1$ is the number of gain molecules in the first excited state. For the used gain medium *i.e.*, RhB, the $N_1/N_g$ have two minimums ~590 and 625 nm as observed (**Figure 4**f). However, the value of $N_1/N_g$ found to be lower at 590 nm than at 625 nm. Thus, it is expected that the gain will be achieved faster for the emission ~590 nm in comparison to that ~625 nm. Therefore, the emission band of RhB dye ~590 nm is observed to narrow down with increase in $P_{in}$. Although, obvious lasing threshold is not noticed in this case due to inefficient localization of photons inside the gain medium. So far, many strategies have been applied for localizing photons inside the gain medium, and hence, the ASE can be converted to RL emission. Likewise, 2D-YP has been introduced into the system as scatterer to achieve RL emission from the gain molecules *i.e*, 2D-YP@RhB. At first, we have measured the $P_{in}$ dependent emission spectra RhB for different number density of 2D-YP scatterer ($N_{sc}$), as shown in **Figure 5**a and **S**2 a-d. The peak intensity and FWHM of the emission spectrum of 2D-YP@RhB ($N_{sc}=16.4 \times 10^9$ nos. ml$^{-1}$) system as a function of $P_{in}$ are shown in **Figure 5**e. In this regard, we have also collected the emission spectra from different direction and the detection angle dependent RL emission spectra and corresponding intensity variation is shown in **Supporting Figure S2** e-f. Clearly, the as designed RL system shown maximum emission intensity at an angle ~40° to the incident pump beam. Therefore, we have fixed the detection of the fiber optic detector ~ 40° with respect to the direction of incidence of the pump beam and rest of the RL experiment is performed thereafter.

The peak emission intensity for the system 2D-YP@RhB is found to be increases slowly upto $P_{in} \approx 65$ mW. However, when the value of $P_{in}$ is larger than 65 mW, the peak intensity of the emission spectrum increases dramatically. Following similar trend, the FWHM of the emission spectrum narrows sharply to ~3 nm when $P_{in}$ exceeds 65 mW. Therefore, we obtain that the pump threshold, *i.e.*, $^{Th}P_{in}$, defined by the sharp narrowing of the emission spectrum and the slope changing of the peak intensity as the $P_{in}$ increases, is about 65 mW. Unlike the bare dye system, sharp RL emission peak ~591 nm appears on the top of SE background with the addition of 2D-YP into the gain medium, and is a signature of ic-RL emission.

Thereafter, the light scattering performance of 2D-YP in generating RL emission has been compared with other materials such as Graphene, 2D-hBN and $TiO_2$, and the results are presented in **Figure 5**b-d, **5**f-h. Henceforth, the RhB solutions containing the Graphene, 2D-Hbn, $TiO_2$ ($N_{sc} \approx 10^9$) are designated as Gr@RhB, 2D-hBN@RhB, and $TiO_2$@RhB, respectively. Notably, the system 2D-hBN@RhB does not show any RL threshold and only ASE behavior is observed with increase in $P_{in}$. Moreover, the slope of the linear variation in the emission intensity *vs.* $P_{in}$ plot is even lower than that of the bare RhB system. Also, FWHM of the ASE in 2D-hBN@RhB is higher than that of the bare RhB (**Figure 5**f). Meanwhile, RL emission is observed when the experiment is performed in the systems $TiO_2$@RhB and Gr@RhB. The value of $^{Th}P_{in}$ and FWHM for $TiO_2$@RhB and Gr@RhB systems is found to be the higher in comparison to that of the 2D-YP@RhB, as summarized in **Figure 5**i.

Here we have used spatial self-phase modulation (SSPM) technique as a tool to track the origin of light scattering for RL emssion from the as designed colloidal disordered mediums. SSPM experiment is performed at $P_{in}$ =80 mW inside a quartz cuvette of path length 0.5 cm. The SSPM patterns for different systems shown in **Figure 5**j are formed due to thermal NLO effect. Clearly number of SSPM rings (*N*) is less for bare RhB solution in comparison to the colloidal systems namely $TiO_2$@RhB and Gr@RhB, respectively.

However, *N* for 2D-hBN@RhB is even less in comparison to that of the bare RhB solution (**Figure 5**j).

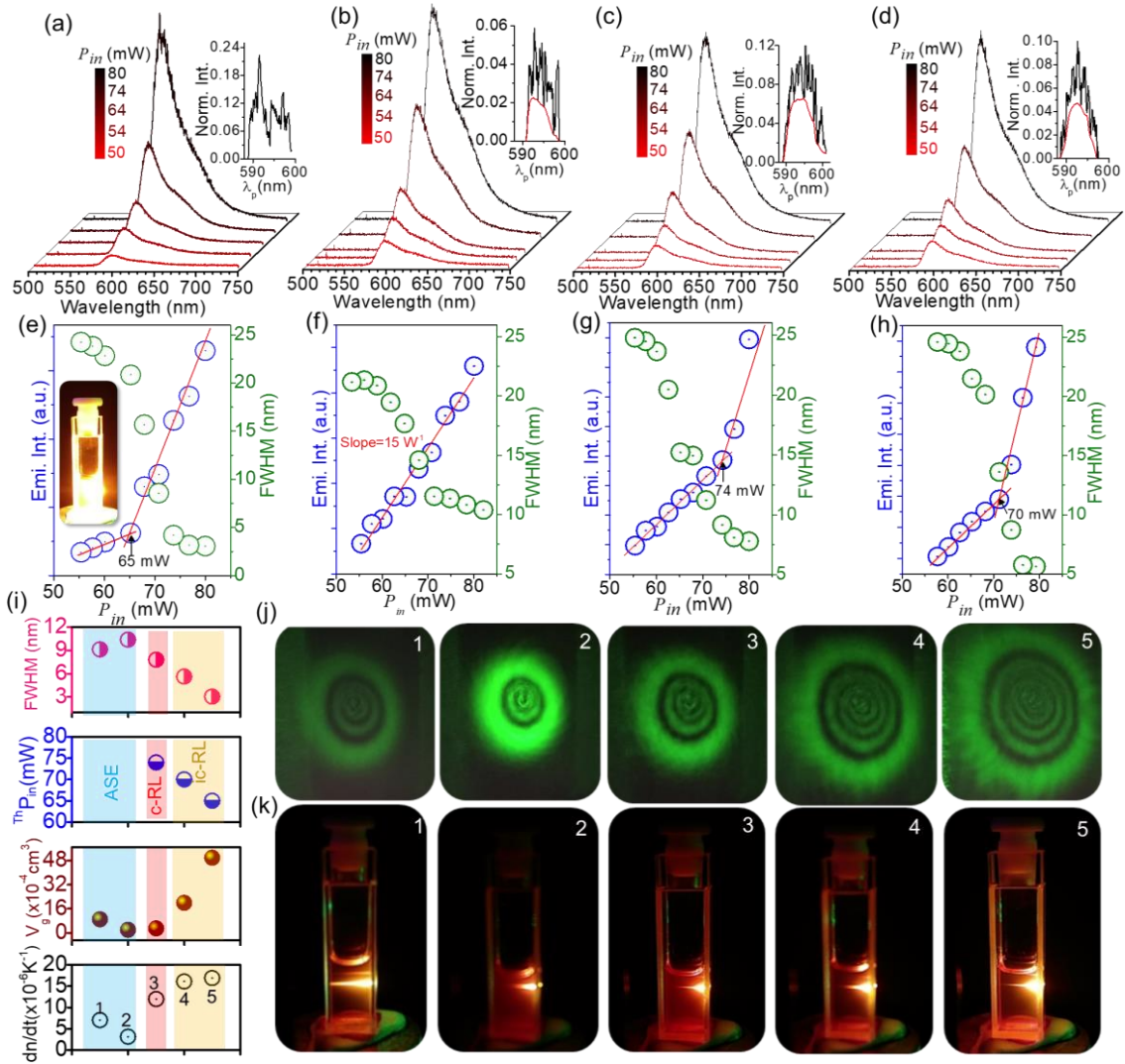

**Figure 5.** Emission spectrum of a) 2D-YP@RhB, b) 2D-hBN@RhB, c) TiO$_2$@RhB and d) Gr@RhB colloidal systems collected at different P$_{in}$. Variation of output emission intensity and FWHM with P$_{in}$ for the system: e) 2D-YP@RhB, f) 2D-hBN@RhB, g) TiO$_2$@RhB and h) Gr@RhB. i) A comparison in the values of dn/dt, V$_g$, $^{Th}$P$_{in}$ and FWHM measured for the disordered systems containing (1) RhB, (2) 2D-hBN@RhB, (3) TiO$_2$@RhB, (4) Gr@RhB, and (5) 2D-YP@RhB. Digital images of j) far-field SSPM patterns, and k) the gain volume during RL experiment performed at P$_{in}$ =80 mW for the system (1) RhB, (2) 2D-hBN@RhB, (3) TiO$_2$@RhB, (4) Gr@RhB and (5) 2D-YP@RhB.

When a laser beam passes through a liquid medium, generation of convection currents takes place due to local heating. Subsequently, a redistribution of local density of the gain molecules as well as the 2D scatterers may take place. This may lead to a formation of temperature dependent change in refractive index within the medium. Generally, thermo-optic coefficient ($dn/dT$) is the parameter which can be used to quantify the change in refractive index with temperature. The values of $dn/dT$ (K$^{-1}$) for each of the systems can be estimated from the corresponding SSPM patterns and by using the relation,

$$\frac{dn}{dT} = \frac{2\pi N K \lambda_p}{1.3 \alpha L_{eff} P_{in}} \left(1 - \varphi \frac{\lambda_p}{\lambda_{em}}\right)^{-1} \qquad (2)$$

Where, $K$ (W m$^{-1}$ K$^{-1}$) is the thermal conductivity of the solution. Generally, $K$ can be considered as thermal conductivity of solvent (IPA), nacre, graphene, hBN and TiO$_2$. In case of liquid systems containing RhB, 2D-YP, graphene, TiO$_2$ and 2D-hBN, respectively. $\lambda_p$= 532 nm is the pump wavelength. $\alpha$ (m$^{-1}$) and $L_{eff}$ (m) are the linear absorption coefficient and effective path length of each the liquid systems, respectively. Here, $\varphi$ (≈0.82) and $\lambda_{em}$ (~592 nm) are the quantum yield and emission wavelength of the gain medium, respectively.

The as estimated value of $dn/dT$ is found to be maximum for 2D-YP@RhB in comparison to that of the other systems. Notably, $dn/dT$ is found to be the lowest for 2D-hBN@RhB. Hence, the largest $dn/dT$ of 2D-YP@RhB system may lead to photon confinement within the random media for achieving RL emission ~591.20 nm at low $^{Th}P_{in}$. In this regard, we have also estimated the value of characteristic length scales such as scattering length ($l_{sc} = 1/N_{sc} \times \sigma_{sc}$), gain length ($l_g = 1/N_g \times \sigma_e$), and transport mean free path ($l_t$) for our disordered system. Where, $d$ and $\sigma_{sc}$ are length (=400 nm) and the scattering cross section of 2D-YP scatterer, $\sigma_e$ is the emission cross-section of the gain medium. In our case, the estimated values of $\sigma_{sc}$, $\sigma_e$, $l_{sc}$, and $l_g$ for the disordered system are found to be $4\times10^{-7}$ cm$^2$, $1.04 \times 10^{-15}$ cm$^2$, 160 μm and 80 μm, respectively. And for Rayleigh scattering limit, ($md/\lambda<1$) $l_{sc} = l_t$.[37] Clearly, the conditions for RL emission in diffusive region, i.e., $\lambda<l_t<L$ and $l_{sc} > l_g$ are satisfied for the as designed disordered active medium of physical length $L=1$ cm. Generally, in diffusive regime, refractive index contrast between scattering and gain medium is reported to be an effective mechanism behind localization and amplification of photons inside the active medium for achieving RL emission.[50-52] Also, in diffusive regime, generation of ic-RL emission may be possible.[53] Whereas, low $dn/dT$ value for the systems TiO$_2$@RhB and Gr@RhB indicates less confinement of photons within the random medium via refractive index contrast. We also investigated the evolution of gain volume ($V_g$) during the RL experiment performed on each kind of random systems by capturing the digital images, as shown in **Figure 5**k, and the values of $V_g$ for different systems are summarized in **Figure 5**i. Owing to the formation of large temperature dependent refractive index gradient perpendicular to the direction of propagation of pump, the value of $V_g$ is also highest for 2D-YP@RhB system. And it causes ic-RL emission at low threshold due to non-resonant intensity feedback. Probability of such non-resonant intensity feedback is less for Gr@RhB and TiO$_2$@RhB systems, as can be understood from the values of $V_g$. However, role of cuvette as a resonator for proving coherent feedback cannot be neglected for the system in which $V_g$ for TiO$_2$@RhB is less. Therefore, generation of multiple coherent lasing modes over the top of the SE background is observed for TiO$_2$@RhB.

Thus, transition from ic-RL to c-RL is possible by regulating the thermo-optical property of the random media. Furthermore, resonant feedback provided by cuvette wall may be responsible for generating coherent RL emission, as we have already observed for

TiO$_2$@RhB system. Also, probability of multiple scattering event inside a disordered system can be controlled by increasing the excitation area. Likewise, we have generated coherent RL emission from 2D-YP@RhB system by (i) altering the diameter of the beam from 0.25 to 0.70 mm, and (ii) by altering the configuration of the sample container.

The evolution of mode patterns on the top of the SE background with increasing beam diameter is shown in **Figure 6**a, and a clear transition from ic-RL to c-RL emission is observed. As the pump diameter increases, possibility of formation of coherent loops due to the multiple scattering events with large dual time enhances successively. [54] Meanwhile, possibility of local heating and, hence the formation of temperature dependent refractive index within the random media for generation of ic-RL emission in diffusive regime is less when excited under pump beam having a comparatively large diameter. A much larger beam diameter causes excessive lowering of $P_{in}$, which is inefficient to excite the RL emission (**Figure 6**a (i-iv)). Thereafter, we have performed the RL experiment inside a cuvette of path length 5 mm and capillary tube of inner diameter 1 mm and the results are summarized in **Figure 6**b-g. Generation of coherent RL emission at $^{Th}P_{in}$ ~62 mW having a highly intense mode ~589 nm is observed from the 2D-YP@RhB in 5 mm cuvette (**Figure 6**b-c). Moreover, the wavelength of the generated RL emission for different configurations has been collected (every 0.5 sec within a time window of 5 min) for statistical analysis, as presented in **Figure 6**d. The wavelength fluctuation ($f_\lambda$) is defined as the ratio of standard deviation ($\sigma_\lambda$) of the emission intensity to the mean emission wavelength ($\lambda_{mean}$). Due to the enhanced possibility of multiple scattering, the value of wavelength fluctuation ($f_\lambda$) is also observed to be increasing from 0.01% to 0.3%, when 2D-YP@RhB is placed inside 5 mm cuvette. Interestingly, 2D-YP@RhB generates RL emission at ~587 nm and at $^{Th}P_{in}$ ~60 mW, when kept inside a capillary tube (**Figure 6**f-g). The effective density of scatterer inside a fixed pump area can have an effect on the c-RL emission characteristic. There will be enough room for the photon to form a closed loop, if the effective number of scatterers inside the pump area is less, and hence the average number of c-RL modes will be high. [55] Likewise, the number of c-RL mode ($M$) is found to be high (~5) when the experiment is preformed inside a 5 mm cuvette, and it may be due to the presence of 2D-YP in less effective density. On the other hand, effective density of 2D-YP will be high when the 2D-YP@RhB is placed inside a capillary tube of inner diameter 1 mm. Therefore, the emitted photons will not get enough room to form closed path inside a fixed pump area, and hence $M$ is observed to be low (down from ~5 to ~2). Also, as observed experimentally the value of $\Delta\lambda$ is high when 2D-YP@RhB is placed inside the capillary tube ($\Delta\lambda \sim 3.0$ nm) in comparison to that of the cuvette of path length 5 mm ($\Delta\lambda \sim 1.8$ nm). Therefore, by keeping 2D-YP@RhB system in two different sample containers, in the present study we have offered a tunable c-RL emission, justified with theoretical background.

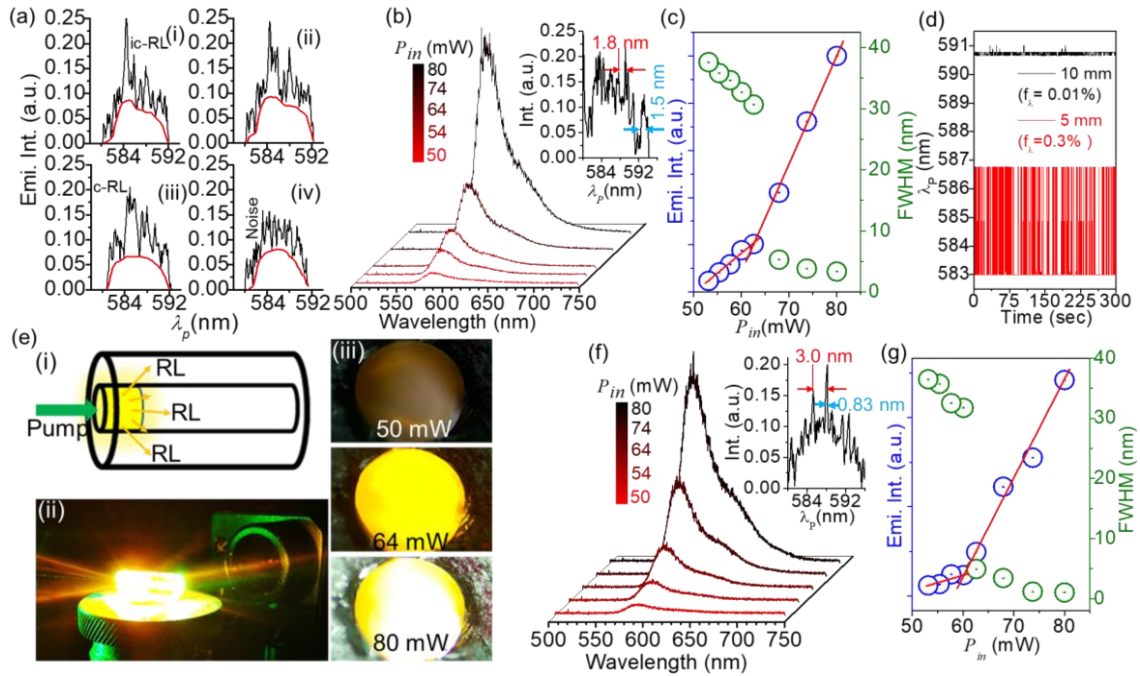

**Figure 6.** a) SE background subtracted emission spectra of 2D-YP@RhB collected at a pump beam diameter of (i) 0.25, (ii) 0.33, (iii) 0.35 and (iv) 0.70 mm and at $P_{in}$ =80 mW. b) Emission spectrum of 2D-YP@RhB colloidal system kept in a cuvette of path length 5 mm and captured at different $P_{in}$. c) Corresponding variation of output emission intensity and FWHM with $P_{in}$. d) The variation in emission wavelength of the observed lasing modes with time carried out for the system at $P_{in}$ of 80 mW, after keeping it inside 10 and 5 mm cuvette. e) (i) Pump configuration of RL experiment performed in capillary tube of inner diameter 1 mm, (ii) Digital image of the as designed RL system in operating condition, (iii) optical microscopic image of active medium (2D-YP@RhB) kept under capillary tube and pumped with 532 nm laser of different pump powers. f) Emission spectrum of 2D-YP@RhB colloidal system kept in capillary tube, and collected at different pump powers. g) The corresponding variation of output emission intensity and FWHM with $P_{in}$.

**CONCLUSIONS**

We have successfully synthesized an uncovational 2D material namely the 2D-YP from bulk south sea yellow pearl by using a simple room temperature LPE approach. The exfoliated 2D-YP is observed to be composed of orthorhombic crystal structure of aragonite with fascinating linear optical and NLO responses. In diffusive regime and under CW pumping, the problem of achiving RL emission from RhB is metigated here by utilizing the structural and optical properties of the passive scatterer *i.e.*, 2D-YP. Notably, unlike other 2D materials, the 2D-YP is reported to have a highest $dn/dT$ ($16 \times 10^{-6}$ K$^{-1}$) within the collioldal disorder system 2D-YP@RhB. Therefore, due to a largest $V_g$ (~$50 \times 10^8$ μm$^3$), the $^{Th}P_{in}$ (FWHM) of the ic-RL in 2D-YP@RhB is reported to be 1.08 (1.84) times lower than the system Gr@RhB. On the other hand, increase in the possibility of multiple scattering wihin the colloidal 2D-YP@RhB may lead to transition of ic-RL to c-RL emission, as we have varified through a phenomological experiment. Therefore, in presence of external feedback, c-RL emssion having a intense lasing mode (linewidth) ~589 nm (~1.5 nm) and 587 nm (0.80 nm) is achived when 2D-YP@RhB is placed inside a 5 nm cuvette and capilary tube of inner diameter 1 mm, respectively. Non-van der Waals 2D materials, such as 2D-YP discussed in this manuscript, will continue to be an important material class for future applications in electronics, optoelectronics, multifunctional coatings, and beyond. As such, the novel 2D-YP based RL is demonstrated as a proof of concept for further exciting applications in the field of non-linear optics and towards developing optoelectronic devices.

## METHODS

### Materials

All materials obtained are of analytical grade and were used without any further purification. Cultured south sea yellow pearls (www.ratnabhandar.com), isopropyl alcohol (IPA, Loba Chemie), de-ionised water were used in the synthesis process. Rhodamine B ($C_{28}H_{31}ClN_2O_3$, Loba Chemie) was used for making dye solutions.

### Synthesis of 2D-YP

A simple liquid exfoliation process was opted for preparing 2D-YP due to its laminar structure. South sea yellow pearls were broken into small pieces using a hammer. These pieces were hand ground into fine powder using an agate mortar and pestle. The material obtained was stored at room temperature for further use. LPE was opted to obtain well-dispersed 2D-YP. The process is shown in **Figure 1**. 5 mg synthesized bulk YP was dissolved in 50 mL iso-propyl alcohol and sonicated for 30-120 min. A probe-sonicator was used for the exfoliation process. To maintain a temperature below 35ºC, a 10 min relaxation time was given to the colloid after each 15 min sonication. After exfoliation, the colloid was stored at room temperature for 24 h to allow settling of non-exfoliated particles. The solution with stable dispersion was then separately stored for further characterization to confirm 2D sheet formation.

**Characterizations**

In order to investigate the microstructural features and physiochemical properties of the synthesized 2D-YP, basic characterization techniques were used where HRTEM images show 2D sheet formation and atomic structure, XRD (PANalytical X'Pert diffractometer) reveals phase and crystallinity, atomic force microscopy (AFM) reveals surface topography and material thickness and a Raman study (WITec, UHTS 300 VIS, Germany) confirms the presence of 2D structure. Linear optical property of the synthesized sample has been measured by using UV-Visible (UV-Vis) spectrophotometer (Analytical) and photoluminescence (PL) spectrofluorimeter (Perkin Elmer LS-55).

**TOC**

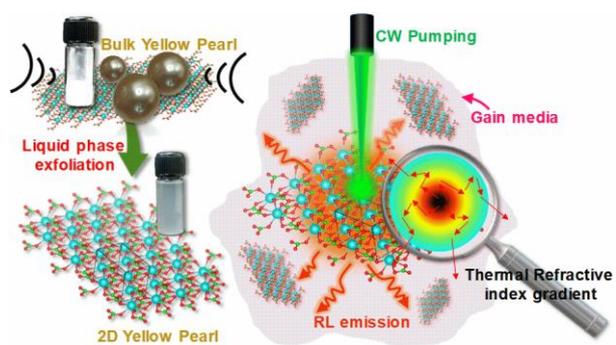